\documentclass[aps,prb,twocolumn,groupedaddress]{revtex4}
\usepackage{amsmath}
\usepackage{graphicx}
\usepackage{ccaption}

\captionnamefont{\small\normalfont}
\captiontitlefont{\small\normalfont} \captionstyle{\flushleftright}
\captiondelim{.} \normalcaption{} \captionwidth{\linewidth}

\begin{document}


\title{Electron Transport in Silicon Nanowires:\\
The Role of Acoustic Phonon Confinement and Surface Roughness Scattering}

\author{E. B.~Ramayya$^{a}$, D.~Vasileska$^{b}$, S. M.~Goodnick$^{b}$, and I. Knezevic$^{a}$}

\affiliation{ $^{a}$Department of Electrical and Computer
Engineering, University of Wisconsin--Madison, Madison, WI 53706,
USA
\\
$^{b}$Department of Electrical Engineering, Fulton School of Engineering, Arizona State
University, Tempe, AZ 85287, USA
}

\date{\today}
\begin{abstract}
We investigate the effects of electron and acoustic-phonon confinement on the low-field electron mobility of thin square silicon nanowires (SiNWs) that are surrounded by SiO$_2$ and gated. We employ a self-consistent Poisson-Schr\"{o}dinger-Monte Carlo solver that accounts for scattering due to acoustic phonons (confined and bulk), intervalley phonons, and the Si/SiO$_2$ surface roughness. The wires considered have cross sections between 3 $\times$ 3 nm$^2$ and 8 $\times$ 8 nm$^2$. For larger wires, as expected, the dependence of the mobility on the transverse field from the gate is pronounced. At low transverse fields, where phonon scattering dominates, scattering from confined acoustic phonons results in about a 10$\%$ decrease of the mobility with respect to the bulk phonon approximation. As the wire cross-section decreases, the electron mobility drops because the detrimental increase in both electron--acoustic phonon and electron--surface roughness scattering rates overshadows the beneficial volume inversion and subband modulation. For wires thinner than 5 $\times$ 5 nm$^2$, surface roughness scattering dominates regardless of the transverse field applied and leads to a monotonic decrease of the electron mobility with decreasing SiNWs cross section. \\

\end{abstract}

\maketitle


\section{Introduction}
Among the emerging devices for the future technology nodes, silicon
nanowires (SiNWs) have attracted much attention among researchers
due to their potential to function as thermoelectric coolers, \cite{BoukaiNAT08,HochbaumNAT08}
logic devices,\cite{CuiJPC00,CuiS01} interconnects,
\cite{LandmanPRL00,SniderNANO07} photodetectors,\cite{ServatiPE07}
as well as biological and chemical sensors.
Although considerable work has recently been done on SiNW, \cite{CuiNL03,KotlyarAPL04,JinJAP07} there is no consensus on these structures' electronic properties.

The low-field electron mobility is one of the most important parameters
that determines the performance of field-effect transistors (FETs), thermoelectric (TE) coolers
and sensors. Although its importance in ultra-short channel MOSFETs
has been debated, \cite{LundstromED02,LundstromIEDMTG03} it
certainly affects the conductivity of the wire interconnects,
the figure of merit of TE coolers and the responsiveness of nanowire sensors.
The study of the electron mobility in SiNWs so far has been inconclusive: Kotlyar {\it et
al}. \cite{KotlyarAPL04} and Jin {\it et
al}. \cite{JinJAP07} have shown that the mobility
in a cylindrical SiNW decreases with decreasing diameter, whereas
the works of Sakaki, \cite{SakakiJJAP80} Cui {\it et al}.,
\cite{CuiNL03} and Koo {\it et al}. \cite{KooNL04} show higher
mobility in SiNWs compared to bulk MOSFETs. The contradiction stems
from two opposing effects that determine the
electron mobility as we move from 2D to 1D structures: one is a
decrease in the density of states (DOS) for scattering
\cite{SakakiJJAP80} that results in reduced scattering rates and
thereby an enhancement in the mobility; the second is an increase in
the so-called electron-phonon wavefunction overlap
\cite{KotlyarAPL04} that results in increased electron-phonon
scattering rates and consequently lower mobility. While important,
these two competing phenomena do not paint a full picture of
low-field transport in SiNWs, in which the effect of spatial
confinement on the scattering due to surface roughness and acoustic
phonons must be addressed.

Surface roughness scattering (SRS) is by far the most important
cause of mobility degradation in conventional MOSFETs at high
transverse fields. Intuitively, one would expect the SRS to be even
more detrimental in SiNWs than conventional MOSFETs because SiNWs
have four Si-SiO$_2$ interfaces, as opposed to one such interface in
conventional MOSFETs. Although recent work has shown that the SRS is in
fact less important in SiNWs than in bulk MOSFETs due to a reduction in
the DOS \cite{WangAPL05} and the onset of volume inversion (redistribution of electrons throughout the silicon channel), \cite{RamayyaIEEEN07} a detailed study of the SRS-limited mobility in cylindrical SiNWs
by Jin {\it et al}. \cite{JinJAP07} shows a rapid monotonic decrease
of mobility for wire diameters smaller than 5 nm.

Scattering due to acoustic phonons is significantly altered in
nanostructures due to the modification of the acoustic phonon
spectrum in them. Extensive work on the effects of acoustic phonon
confinement in III-V based nanostructures
\cite{MduttaIJHSES98,BalandinJNN05} shows a lower acoustic phonon
group velocity, \cite{PokatilovSM03,PokatilovPRB05} lower thermal
conductivity, \cite{LuJAP03,LuJAP06,ChenJHT05} and increased
acoustic phonon scattering rates \cite{SvizhenkoJPCM98,YuPRB94} in
nanoscale devices compared to their bulk counterparts. As for
silicon nanostructures, there is experimental evidence of the
acoustic phonons confinement in nanomembranes, \cite{TorresPSSC04}
while recent works on SOI MOSFETs \cite{GamizSOI05,DonettiJAP06,DonettiAPL06} and SiNWs \cite{BuinNL08}
also show that the bulk phonon (linear dispersion) approximation underestimates the scattering rates.

In this work, we calculate the electron mobility of gated square SiNWs by using a self-consistent Schr\"{o}dinger-Poisson-Monte Carlo simulator and  accounting for electron scattering due to
acoustic phonons (confined and bulk), intervalley phonons, and imperfections at
the Si/SiO$_2$ interface. The wires considered have cross sections between 3 $\times$ 3 nm$^2$ and 8 $\times$ 8 nm$^2$.
Bulk-silicon effective mass parameters are used in the calculation of the scattering rates.
The confined acoustic phonon spectrum is obtained by using the \emph{xyz} algorithm
\cite{VisscherJASA91,NishiguchiJPCM97} and the unscreened SRS is
modeled using modified Ando's model \cite{AndoRMP82} that accounts for the finite thickness of the silicon layer.
For larger wires, as expected, the dependence of the mobility on the transverse field from the gate is pronounced;  at low transverse fields, where phonon scattering dominates, scattering from confined acoustic phonons results in about a 10$\%$-decrease of the mobility with respect to the bulk phonon approximation. As the wire cross section decreases, the electron mobility drops because the detrimental increase in both electron--acoustic phonon and electron-surface roughness scattering rates overshadows the beneficial effects of volume inversion (redistribution of electrons in real space) \cite{BalestraEDL87,RamayyaIEEEN07} and subband modulation (redistribution of electrons among different subbands) \cite{TakagiJJAP98,GamizAPL04,UchidaJAP07}. For wires thinner than 5 $\times$ 5 nm$^2$, surface roughness scattering dominates regardless of the transverse field applied and leads to a monotonic decrease of the electron mobility with decreasing SiNWs cross-section.

The electronic bandstructure in SiNWs is altered from that of bulk silicon due to the Brillouin zone folding. \cite{WangED05, ZhengED05, BuinNL08} The degeneracy of the conduction band minima in SiNWs is split; $\Delta_4$ valleys (four degenerate $\Delta$ valleys whose long axis is perpendicular to the SiNW axis) are found to be at the $\Gamma$ point and the $\Delta_2$ valleys (two degenerate $\Delta$ valleys whose long axis is parallel to the SiNW) are at $k_x=\pm0.37\pi/a$. The bandstructure modification will certainly affect the intervalley scattering rates in SiNWs, but deformation potentials and phonon energies needed to describe the intervalley scattering are still only available for bulk silicon. Therefore, in order to be consistent, we do not employ the exact nanowire bandstructure, but rather rely on the bulk Si bandstucture and then solve the two-dimensional Schr\"{o}dinger equation within the the envelope function and effective mass framework, with the bulk Si effective mass   parameters. This approximation was proven adequate down to the 3 nm wire diameter.\cite{WangED05} The splitting of the sixfold degenerate $\Delta$ valleys of bulk silicon into $\Delta_2$ and $\Delta_4$ upon confinement is automatically accounted for by including the anisotropy of the electron effective mass (see Sec. \ref{SubMod}).

This paper is organized as follows: Section \ref{DevStr} describes
the device structure used in this study and the components of the
simulator developed to calculate the mobility. A description of the
acoustic phonon spectrum calculation in SiNWs is given towards
the end of Section \ref{DevStr}. In Section \ref{SimRes}, we
emphasize the importance of accounting for the acoustic phonon
confinement when calculating the mobility in SiNWs and then present
the results for the variation of mobility in square SiNWs with
increasing spatial confinement. We conclude this paper with a brief
summary of the findings of our work (Section \ref{Concl}) and a
detailed derivation of the scattering rates due to acoustic phonons,
intervalley phonons, and surface roughness (Appendices
\ref{phScatDeriv} and \ref{SRSDeriv}).

\begin{figure}
\includegraphics[width=2.5in]{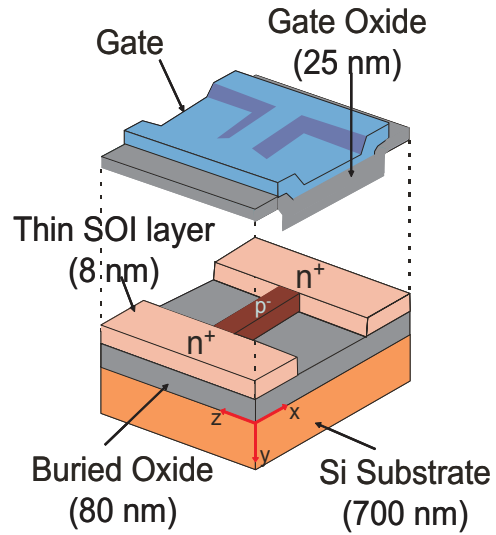}
\caption{$\quad$(Color online) Schematic of the simulated 8$\times$8 nm$^2$ SiNW on ultrathin SOI.}\label{DevStrPotY}
\end{figure}

\section{Mobility Calculation}\label{DevStr}
\subsection{Device Structure and Simulator Components}

A schematic of the device considered in our study is shown in the
top panel of Fig. \ref{DevStrPotY}. It is a modified version of the
ultra-thin, ultra-narrow SOI MOSFET that was originally proposed by
Majima {\it et al}.\cite{MajimaEDL00} The thickness of the gate oxide,
buried oxide, and bottom silicon substrate are 25 nm, 80 nm, and 700 nm, respectively.
The transverse dimensions of the silicon channel are varied from 8 $\times$ 8 nm$^2$ to 3
 $\times $3 nm$^2$. For all the device cross sections considered, the
width of the oxide on both sides of the channel is 200 nm, the
channel is doped to $3\times 10^{15}$ cm$^{-3}$, and the channel is
assumed to be homogeneous and infinitely long.

The simulator developed to calculate the electron mobility has two
components: the first is a self-consistent 2D Poisson-2D
Schr\"{o}dinger solver and the second is a Monte Carlo transport
kernel. The former is used to calculate the electronic states and
the self-consistent potential distribution along the cross section
of the wire and the latter simulates the transport along the wire
axis. The finite barrier at the Si/SiO$_2$ interface results in the
electron wavefunction penetration through the interface and into the
oxide. The wavefunction penetration is accounted for by including a
few mesh points in the oxide while solving the Schr\"{o}dinger
equation. ARPACK package \cite{ARPACK} was used to solve the 2D
Schr\"{o}dinger equation and the successive over-relaxation (SOR) method
was used to solve the 2D Poisson equation. The convergence of the coupled
Schr\"{o}dinger-Poisson solver is found to be faster when the Poisson equation is solved by using the SOR method than when the incomplete lower-upper (ILU) decomposition method is used.

The Monte Carlo transport kernel is used to simulate the electron
transport along the axis of the wire under the influence of the
confining potential in the transverse directions and a very small
lateral electric field along the channel. The long wire
approximation implies that the transport is diffusive (the length
exceeds the carrier mean free path), and therefore justifies the use
of the Monte Carlo method \cite{JacoboniRMP83,FischettiPRB93} to
simulate electron transport. Electrons are initialized such that
their average kinetic energy is $(1/2)K_BT$ (thermal energy for 1D)
and are distributed among different subbands in accordance with the
equilibrium distribution of the states obtained from the Poisson-Schr\"{o}dinger solver. Since the electrons are confined in two transverse directions, they are only scattered in either the forward or the backward direction;
consequently, just the carrier momentum along the length of
the wire needs to be updated after each scattering event. Mobility
is calculated from the ensemble average of the electron
velocities.\cite{JacoboniRMP83}

\subsection{Scattering due to Bulk Acoustic Phonons, Intervalley Phonons, and Surface Roughness}

Phonon scattering and the SRS are considered in this work. The SRS
was modeled using Ando's model, \cite{AndoRMP82} intervalley
scattering was calculated using bulk phonon approximation, and the
intravalley acoustic phonons were treated in both the bulk-mode and
confined-mode approximations. Since the wire is very lightly doped,
the effect of impurity scattering was not included. Nonparabolic
band model for silicon, with the nonparabolicity factor $\alpha=0.5
eV^{-1}$, was used in the calculation of scattering rates.
A detailed derivation of the 1D scattering rates is given in
Appendices \ref{phScatDeriv} (phonon scattering) and \ref{SRSDeriv}
(SRS). Here, for brevity, only the final expressions for the
scattering rates are given.

For an electron with an initial lateral wavevector $k_x$ and
parabolic kinetic energy $\mathcal{E}_{k_x}=\hbar^2k^2_x/(2m^*)$ in
subband $n$ [with subband energy $\mathcal{E}_n$ and electron
wavefunction $\psi_n(y,z)$], scattered to subband $m$ [with subband
energy $\mathcal{E}_m$ and electron wavefunction $\psi_m(y,z)$], the
final kinetic energy $\mathcal{E}_f$ is given by
\begin{equation}\label{Ef}
\mathcal{E}_f=\mathcal{E}_n-\mathcal{E}_m+\frac{\sqrt{1+4\alpha\mathcal{E}_{k_x}}-1}{2\alpha}+\hbar\omega,
\end{equation}
where $\hbar\omega=0$ for elastic (bulk intravalley acoustic phonon
and surface roughness) scattering, $\hbar\omega=\pm\hbar\omega_0$
for the absorption/emission of an approximately dispersionless
intervalley phonon of energy $\hbar\omega_0$, while in the case of
confined acoustic phonons (below) the full phonon subband dispersion
is incorporated.

The intravalley acoustic phonon scattering rate due to bulk acoustic
phonons is given by
\begin{equation}\label{AcScatRate}
\Gamma^{ac}_{nm}(k_x)=\frac{\Xi^2_{ac}k_BT\sqrt{2m^*}}{\hbar^2\rho\upsilon^2}\
\mathcal{D}_{nm}\frac{(1+2\alpha\mathcal{E}_f)}{\sqrt{\mathcal{E}_f(1+\alpha\mathcal{E}_f)}}\
\Theta(\mathcal{E}_f),
\end{equation}
where $\Xi_{ac}$ is the acoustic deformation potential, $\rho$ is
the crystal density, $v$ is the sound velocity, and $\Theta$ is the
Heaviside step-function. $\mathcal{D}_{nm}$ represents the overlap
integral associated with the electron-phonon interaction (the
so-called electron-phonon wavefunction integral\cite{KotlyarAPL04}),
and is given by
\begin{equation}\label{OverlapPh}
\mathcal{D}_{nm}=\iint|\psi_n(y,z)|^2|\psi_m(y,z)|^2\,dy\,dz.
\end{equation}

The intervalley phonon scattering (mediated by short wavelength
acoustic and optical phonons) rate is given by
\begin{equation}\label{IvScatRate}
\begin{aligned}
\Gamma^{iv}_{nm}(k_x)=\frac{\Xi^2_{iv}\sqrt{m^*}}{\sqrt{2}\hbar\rho\omega_0}\
&\left(N_{0}+\frac{1}{2}\mp\frac{1}{2}\right)\
\mathcal{D}_{nm} \\
&\times\frac{(1+2\alpha\mathcal{E}_f)}{\sqrt{\mathcal{E}_f(1+\alpha\mathcal{E}_f)}}\
\Theta(\mathcal{E}_f),
\end{aligned}
\end{equation}
where $\Xi_{iv}$ is the intervalley deformation potential, and
$\mathcal{D}_{nm}$ is defined in (\ref{OverlapPh}). The
approximation of dispersionless bulk phonons of energy
$\hbar\omega_0$ was adopted to describe an average phonon with
wavevector near the edge of the Brillouin zone and
$N_{0}=\left[{\exp(\hbar\omega_0/k_BT)-1}\right]^{-1}$ is their
average number at temperature $T$.

Assuming exponentially correlated surface roughness
\cite{GoodnickPRB85} and incorporating the electron wavefunction
deformation due to the interface roughness using Ando's model,
\cite{AndoRMP82} the unscreened SRS rate is given by
\begin{equation}\label{SRScatRate}
\begin{aligned}
\Gamma^{sr}_{nm}(k_x,\pm)=\frac{2\sqrt{m^*}e^2}{\hbar^2}&\frac{\Delta^2\Lambda}{2+(q^{\pm}_x)^2\Lambda^2}|\mathcal{F}_{nm}|^2
\\
\times
&\frac{(1+2\alpha\mathcal{E}_f)}{\sqrt{\mathcal{E}_f(1+\alpha\mathcal{E}_f)}}\
\Theta(\mathcal{E}_f),
\end{aligned}
\end{equation}
where $\Delta$ and $\Lambda$ are the r.m.s. height and the
correlation length of the fluctuations at the Si-SiO$_2$ interface,
respectively. $q_x^{\pm}=k_x{\pm}k_x'$ is the difference between the
initial ($k_x$) and the final ($k_x'$) electron wavevectors and the
top (bottom) sign is for backward (forward) scattering. The SRS
overlap integral in Eq. (\ref{SRScatRate}) due to the top interface for a silicon body thickness of $t_y$ is given by
\begin{eqnarray}\label{OverlapSRy}
\mathcal{F}_{nm}&=&\iint
dy\,dz\left[-\frac{{\hbar}^2}{{e}{t_y}{m_y}}{\psi_m(y,z)}\frac{\partial^2{\psi_n(y,z)}}{\partial{y^2}}
\right. \nonumber\\
&+&\left.\psi_n(y,z)\varepsilon_y(y,z)\left(1-\frac{y}{t_y}\right)\psi_m(y,z) \right.\\
&+&\left.\psi_n(y,z)\left(\frac{\mathcal{E}_m-\mathcal{E}_n}{e}\right)\left(1-\frac{y}{t_y}\right)\frac{\partial{\psi_m(y,z)}}{\partial{y}}\
\right].\nonumber
\end{eqnarray}

The SRS overlap integral was derived assuming the interfaces to be uncorrelated. For the bottom interface, the integration should be performed from the bottom interface to the top interface and the integral for the side interfaces can be obtained by interchanging $y$ and $z$ in Eq. (\ref{OverlapSRy}). The first term in Eq. (\ref{OverlapSRy}) is the confinement-induced part of the SRS and it increases with decreasing wire cross section. This term does not depend on the position of the electrons in the channel and hence results in high SRS even at low transverse fields from the gate. The second and third terms in Eq. (\ref{OverlapSRy}) depend on the average distance of electrons from the interface, so they contribute to the SRS only at high transverse fields from the gate. (In Sec. \ref{SimRes}, for example, we will see that the major contribution to the SRS in wires thinner than 5 $\times$ 5 nm$^2$ comes from the confinement-induced term in Eq. (\ref{OverlapSRy}), and it increases rapidly with decreasing wire cross section.)

Scattering rates given by Eqs. (\ref{AcScatRate})--(\ref{SRScatRate}) are calculated using the
wavefunctions and potential obtained from the self-consistent
Poisson-Schr\"{o}dinger solver. For this work, the parameters used for
calculating the intervalley scattering were taken from Ref. \onlinecite{TakagiJJAP98},
the acoustic deformation potential was taken from Ref. \onlinecite{BuinNL08}, and
$\Delta$ = 0.3 nm and $\Lambda$ = 2.5 nm were used to characterize
the SRS due to each of the four interfaces. The SRS parameters were obtained by fitting the mobility of an 8 $\times$ 8 nm$^2$ SiNW in the high transverse field region (where the SRS dominates) with the corresponding mobility observed in ultra-thin SOI of similar thickness.\cite{UchidaJAP07}

\subsection{Acoustic Phonon Confinement}\label{AcPhConf}

In ultra-thin and ultra-narrow structures, the acoustic phonon
spectrum is modified due to a mismatch of the sound velocities and
mass densities between the active layer and the surrounding
material, \cite{PokatilovPRB05,PokatilovSM05} in our case -- silicon
and SiO$_2$. This modification in the acoustic phonon spectrum
becomes more pronounced as the dimensions of the active layer become
smaller than the phonon mean free path, which is around 300 nm in
silicon.\cite{JuAPL99} Pokatilov {\it et al.}\cite{PokatilovPRB05}
have shown that the modification in the acoustic phonon dispersion
in nanowires can be characterized by acoustic impedance
$\zeta={\rho}V_s$, where $\rho$ and $V_s$ are the mass density and
sound velocity in the material, respectively. By considering
materials with different $\zeta$, Pokatilov {\it et al.}\cite{PokatilovPRB05} have shown that the acoustic phonon group velocity in the active layer is
reduced when an acoustically soft (smaller $\zeta$) material surrounds
an active layer made of acoustically hard (higher $\zeta$) material.
Since Si is acoustically harder than SiO$_2$, the acoustic phonon
group velocity in SiNWs with SiO$_2$ barriers decreases and
results in an increased acoustic phonon scattering rate
[see Eq. (\ref{AcScatRate})].

\subsubsection{Confined Acoustic Phonon Dispersion}

\begin{figure} [h]
\centering \centering
\includegraphics[width=2.5in]{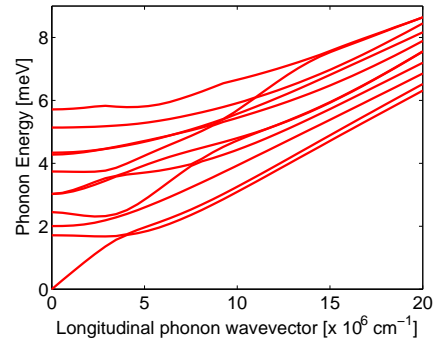}
\caption{$\quad$(Color online) Confined acoustic phonon dispersion (dilatational mode) calculated using the \emph{xyz} algorithm \cite{NishiguchiJPCM97} for an 8 $\times$ 8
nm$^2$ SiNW. Only the lowest 10 phononic subbands are shown.
Dispersion in the first one third of the first Brillouin zone is shown for
clarity.}\label{AcPhDisp8nm}
\end{figure}

The first step in accounting for the acoustic phonon confinement in
mobility calculation is to calculate the modified acoustic phonon
dispersion. Using the adiabatic bond charge model \cite{WeberPRB77}
(microscopic calculation, accurate but computationally involved),
Hepplestone and Srivastava have shown the validity of the elastic
continuum model (macroscopic calculation, less accurate but easier
to implement) for wire dimensions greater than 2.5 nm.
\cite{HepplestonePSSC04} Hence, in this work we have used the
elastic continuum model to calculate the modified phonon spectrum. Most of
the previous studies of acoustic phonon confinement in nanowires
have used approximate hybrid modes proposed by Morse
\cite{MorseJASA50} (valid for wires with thickness much smaller than the width) to calculate the dispersion spectrum.  Nishiguchi
{\it et al.} \cite{NishiguchiJPCM97} calculated the dispersion
spectrum using the \emph{xyz} algorithm \cite{VisscherJASA91} and found
that the Morse formalism is valid only for the lowest phonon
subband. Since one acoustic phonon subband is certainly not enough to
accurately describe scattering with electrons, in this work, we have
used Nishiguchi {\it et al.}'s approach to calculate the acoustic
phonon dispersion, although it is computationally intensive. The
basis functions used to expand the phonon mode displacements in
Nishiguchi {\it et al.}'s approach are powers of Cartesian
coordinates, and the number of basis functions required to fully describe
the modes depends on the number of modes required. For
an 8 $\times$ 8 nm$^2$ SiNW, we have found that the lowest 35
phononic subbands are enough to calculate the scattering rate. Also,
we found that about 176 basis functions are sufficient to fully
describe the displacement of those 35 phononic modes. The number of
phononic bands required decreases with a decrease in the wire cross section.

Two types of boundary conditions are often used to calculate the
acoustic phonon spectrum in nanostructures: a) The free-standing
boundary condition (FSBC) assumes that all the surfaces are free, so
normal components of the stress tensor vanish at the surfaces, and
b) The clamped-surface boundary condition (CSBC) assumes that the
surfaces are rigidly fixed, so the displacement of phonon modes is
zero at the surfaces. Generally, the CSBC (FSBC) results in higher
(lower) phonon group velocity than the bulk case.
\cite{DonettiJAP06} For the wires considered in this work, neither
of these boundary conditions holds exactly, since these wires are actually
embedded in the SiO$_2$. Ideally, one needs to solve the elastic
continuum equation, taking into account the continuity of
displacement and stress at all Si-SiO$_2$ interfaces, and then apply
the boundary conditions at the outer surfaces. But, this is almost
numerically impossible for the structure considered because it would
be equivalent to solving the 2D Schr\"{o}dinger equation in a device
with the cross section of about 800 $\times$ 400 nm$^2$, three (five
if the metal-Si interface is included) interfaces along the depth, and
two interfaces along the width. Donetti {\it et al.}, \cite{DonettiJAP06} in their work
on a SiO$_2$/Si/SiO$_2$ sandwich structure, considered continuity of the
displacement and stress at the interfaces to calculate the phonon
dispersion.  They found it to be close to the
results from the FSBC. Therfore, in this work, we have used the FSBC to
calculate the acoustic phonon spectrum of SiNWs. Fig. \ref{AcPhDisp8nm} shows
the calculated acoustic phonon dispersion of the lowest 10 dilatational modes for an 8 $\times$ 8 nm$^2$ SiNW.
Apart from these dilatational modes, depending on the rotational symmetry of the confined acoustic phonon displacement, there are two sets of flexural modes and one set of torsional modes in SiNWs. A detailed description of the symmetry of all these phonon modes can be found in Ref. \onlinecite{NishiguchiJPCM97}.

\subsubsection{Scattering due to Confined Acoustic Phonons}\label{ConfAcImp}

\begin{figure} [h]
\centering \centering
\includegraphics[width=2.5in]{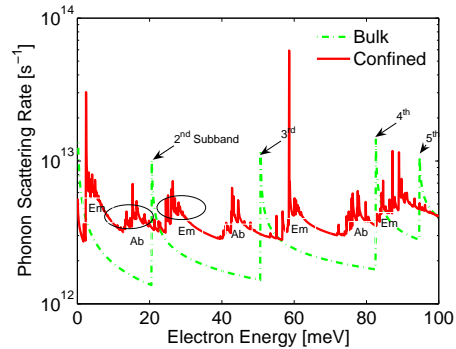}
\caption{  $\quad$(Color online)  Electron -- acoustic phonon scattering rate for the lowest electron subband of an 8 $\times$ 8
nm$^2$ SiNW at the channel sheet density of $N_s=8.1\times10^{11}$
cm$^{-2}$, calculated assuming the bulk (dash-dot green line) and confined phonons (solid red line). The electron-bulk acoustic phonon intersubband spikes are at around 20 meV, 52 meV, 85 meV, and 95 meV, and they correspond to the electron scattering from the lowest subband to the 2$^{nd}$, 3$^{rd}$, 4$^{th}$, and the 5$^{th}$ subbands, respectively. To the left/right of
each intersubband scattering spike that corresponds to the bulk phonon approximation (dash-dot green) are two groups of
small spikes (solid red) that correspond to absorption ("Ab")/emission ("Em") of confined phonons from different phonon subbands.} \label{ConfAcBulkAcScat}
\end{figure}

\begin{figure} [h]
\centering \centering
\includegraphics[width=2.5in]{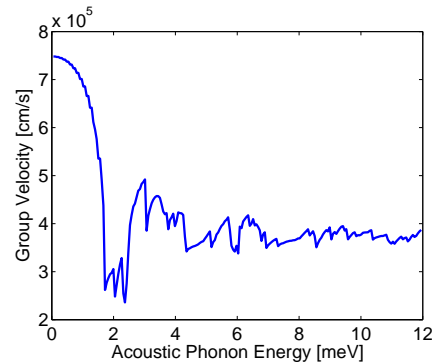}
\caption{  $\quad$(Color online) Group velocity of dilatational modes for an 8$\times$8 nm$^2$ SiNW. On average, acoustic phonon group velocity is reduced to less than 50$\%$ of the bulk value (9.13 $\times$10$^5$ cm/s).} \label{fig:group velocity}
\end{figure}

The modification of the acoustic phonon dispersion due to
confinement, shown in Fig. \ref{AcPhDisp8nm}, implies that the
linear dispersion and elastic scattering approximation can no longer
be used in calculating the scattering rate. The modified scattering
rate which takes into account confined acoustic phonon modes is
given by
\begin{eqnarray}\label{ConfAcPhScatRate}
\Gamma_{nm}^{ac}(k_x)&=&\frac{\Xi_{ac}^{2}}{2WH}\sum_{J}\sum_{i=1,2}
\left(N_{Jq_{x_i}}+\frac{1}{2}{\pm}\frac{1}{2}\right)\left|\alpha_{J}\right|^{2}\nonumber\\
&\times
&\frac{\left|\mathcal{L}_{nm}(J,q_{x_i})\right|^{2}}{\left|g'(q_{x_i})\right|} ,
\end{eqnarray}
where $q_x$ is the lateral wavevector of the acoustic phonon,
$g(q_{x})=E-E'{\mp}\hbar\omega_{J}(q_x)$, $q_{x_1}$ and $q_{x_2}$
are the two possible roots of $g(q_{x})=0$, and $g'(q_{x_1})$ and
$g'(q_{x_2})$ are the derivatives of $g(q_{x})$ with respect to
$q_{x}$ evaluated at $q_{x_1}$ and $q_{x_2}$, respectively. Index
$J$ stands for the different acoustic phonon modes and $N_{Jq_{x}}$
is the number of acoustic phonons of energy $\hbar\omega_{J}(q_x)$.
Overlap integral $\mathcal{L}_{nm}(J,q_{x})$ and the total energy of the
electron before ($E$) and after ($E'$) scattering are defined in
Appendix \ref{ConfAcPhDeriv}.

For intrasubband transitions, only dilatational modes are important, because for all other modes the overlap integral $\mathcal{L}_{nm}(J,q_{x})$ in  Eq.(\ref{ConfAcPhScatRate}) vanishes due to symmetry. In intersubband transitions, all the four sets of acoustic phonon modes are included in the calculation of the electron -- confined acoustic phonon scattering rates, but the dominant contribution to the scattering rate comes from the dilatational modes.

Fig. \ref{ConfAcBulkAcScat} shows the intrasubband electron--acoustic phonon scattering rate for the lowest electron subband, calculated using both the bulk-mode and confined-mode approximations. When calculating
the electron--bulk acoustic phonon scattering rates, acoustic phonon dispersion is assumed to be linear, $\omega_q = V_s q$, where $V_s$ is the sound velocity, as before. The resulting scattering rate, in the elastic and equipartition approximations (see Appendix A1), is proportional to the final electron density of
states, and has the characteristic 1D density-of-states peaks (dot-dashed green line) whenever the
electron energy becomes sufficient to scatter into the next subband. In the case of confined acoustic phonons, as seen in Fig. \ref{AcPhDisp8nm}, the elastic approximation for electron-phonon scattering
no longer holds, and neither does the linear dispersion at small wavevectors (except for the lowest phonon subband). Still, one can speak of a group velocity associated with a collection of phononic subbands. The average group velocity accounting for the non-uniform energy
gap between different phonon modes is shown in Fig. \ref{fig:group velocity}. The average group velocity is close to the bulk value for very small phonon energies, but asymptotically reaches a constant
value (less than 50$\%$ of its bulk value) at high phonon energies. Since the scattering rate due to confined phonon subbands is inversely proportional to their group velocity, on average,
the confined acoustic phonon scattering rate is about two times the acoustic scattering rate calculated using bulk phonons (Fig. \ref{ConfAcBulkAcScat}). Moreover, each of the bulk-phonon-scattering
intersubband peaks in Fig. \ref{ConfAcBulkAcScat} (obtained in the elastic approximation, so the absorption and
emission rate peaks coincide) splits into two groups of peaks when confined
phonons are considered: confined phonons can generally not be treated as elastic, hence
there is a group of small peaks due to confined-phonon absorption below each bulk-phonon-
scattering peak associated with a given electron subband, and a group of peaks due to confined-phonon emission above the bulk-phonon-scattering peak.


\section{Electron Mobility in Silicon Nanowires -- Simulation Results}\label{SimRes}

In our previous work, \cite{RamayyaIEEEN07} we examined the
variation of the transverse field-dependent mobility with decreasing
channel width in 8 nm thick rectangular SiNW of different widths and observed two that (i) the mobility at low-to-moderate transverse
fields, limited by phonon scattering, decreases with decreasing
wire width, and (ii) the mobility at high transverse fields, which
is dominated by the SRS, increases with decreasing wire width. The
former is due to the increase in the electron-phonon wavefunction
overlap [Eq. (\ref{OverlapPh})] with decreasing wire width, and the
latter is due to the onset of volume inversion. In this section, we
first emphasize the importance of acoustic phonon confinement in
SiNWs, and then vary the cross section of the wire to investigate the
effect of increasing spatial confinement on electron mobility.

\begin{figure} [h]
\centering \centering
\includegraphics[width=2.5in]{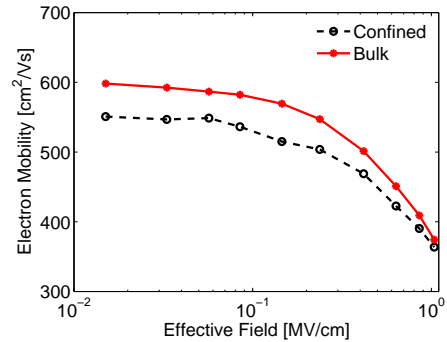}
\caption{  $\quad$(Color online) Variation of the field-dependent mobility for an 8
$\times$ 8 nm$^2$ SiNW assuming bulk acoustic phonons (solid line) and confined
acoustic phonons (dashed line).} \label{MuConfBulk}
\end{figure}

The electron mobility in an 8 $\times$ 8 nm$^2$ wire, with and without phonon confinement, is shown in Fig. \ref{MuConfBulk}. In the low-transverse-field region, the mobility calculated with confined acoustic phonons is about 10 $\%$ lower than that obtained with bulk phonons. This
clearly indicates that confined acoustic phonons need to be properly
included in the study of electrical transport in SiNWs. The mobility values for 8 $\times$ 8 nm$^2$ are very close to the experimentally observed mobility in ultra-thin SOI of similar thickness.\cite{UchidaJAP07}
\textit{In the remainder of the article, we will always assume confined acoustic phonons. }

\begin{figure} [h]
\centering \centering
\includegraphics[width=2.5in]{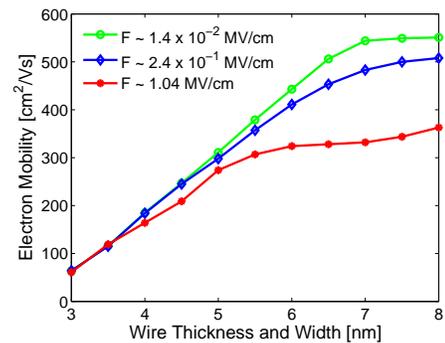}
\caption{  $\quad$(Color online) Variation of the electron mobility with SiNW cross
section at three different transverse fields.} \label{MuVsCross}
\end{figure}

To determine the cross sectional dependence of the electron mobility in SiNWs and to understand the confinement effects on the spatial and \emph{k}-space distribution of electrons, the cross section of the wire was varied from 8 $\times$ 8 nm$^2$ to 3 $\times$ 3 nm$^2$. The variations of the electron mobility with decreasing wire cross section at a low (1.4 $\times10^{-2}$ MV/cm), moderate (2.4 $\times10^{-1}$ MV/cm) and high (1.04 MV/cm) transverse field are plotted in Fig. \ref{MuVsCross}. In the following, we will see that a complex interplay of several competing physical mechanisms is responsible for the electron mobility behavior observed in Fig. \ref{MuVsCross}.

\subsection{Subband Modulation}\label{SubMod}

One of the most important factors that determines the energy and
occupation probability of a subband in each of the $\Delta_6$ valleys
(equivalent in bulk silicon) is the effective mass in the direction of confinement.
For the SiNWs considered, the confinement is
along the $y$ and $z$ directions and the electrons are allowed to
move freely in the $x$ direction; consequently, the conductivity
effective mass for the valley pairs with minima on the $x$, $y$, and
$z$ axes are $m_l, m_t$, and $m_t$, respectively, while their
subband energies are roughly proportional to $1/\sqrt{m_t^2}$, $1/\sqrt{m_t
m_l}$, and $1/\sqrt{m_t m_l}$, respectively. Since $m_t < m_l$, the
subbands in the valley pair along $x$ are higher in energy than
those in the valley pairs along $y$ and $z$. So the subbands split
into those originating from the twofold degenerate $\Delta_2$ (the
valley pair along $x$) and those originating from the fourfold
degenerate $\Delta_4$ valleys (the valley pairs along $y$ and $z$).

\begin{figure} [h]
\centering \centering
\includegraphics[width=2.5in]{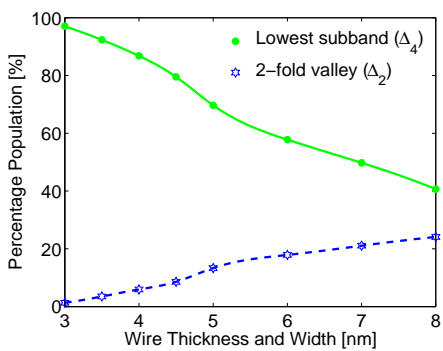}
\caption{  $\quad$(Color online) Variation of the electron population at a low
transverse field (1.4 $\times10^{-2}$ MV/cm). The solid line shows
the total population of electrons in the lowest subband in the $\Delta_4$ valleys and the dashed line shows population of
the $\Delta_2$ valley pair with varying spatial confinement.}
\label{DistVsCross}
\end{figure}

Upon increasing spatial confinement by decreasing the wire cross
section, the subbands in different valleys are pushed
higher up in energy, and consequently only a few of the lowest
subbands in each of the valley pairs get populated with electrons.
Fig. \ref{DistVsCross} shows the depopulation of the higher $\Delta_2$
valley subbands with increasing spatial confinement: since the
lowest subbands in the $\Delta_4$ valleys are lower in energy than
those in the $\Delta_2$ valleys, under extreme confinement
$\Delta_2$ subbands get completely depopulated, and only the lowest
$\Delta_4$ subbands are populated. Splitting of the valley degeneracy and  modification of the subband energies in different valley pairs due to spatial
confinement, followed by depopulation of the higher subbands, are together  termed \textit{subband modulation}.\cite{TakagiJJAP98} Subband modulation enhances the electron mobility becauses it suppresses intersubband and intervalley scattering, as shown for ultrathin-body SOI MOSFETs.\cite{UchidaJAP07} In our previous work on SiNWs,\cite{RamayyaJCE08} we also observed a small enhancement in mobility for wires of cross section around 4 $\times$ 4 nm$^2$, but in that study we did not consider the confinement-induced term in SRS. However, this term in the SRS increases rapidly for wire cross section below 5 $\times$ 5 nm$^2$ and suppresses the beneficial effect that subband modulation has on the electron mobility in ultra-small SiNWs. Indeed, the confinement-induced SRS term has been shown to be dominant in determining the SRS in small cylindrical  SiNWs.\cite{JinJAP07}

\subsection{Volume Inversion}\label{VolInv}

\begin{figure} [h]
\centering \centering
\includegraphics[width=2.5in]{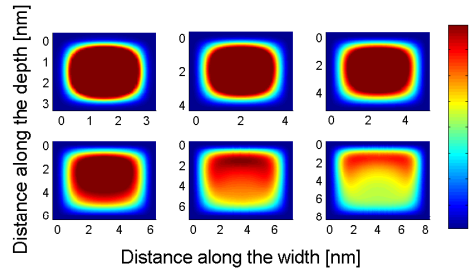}
\caption{  $\quad$(Color online) Electron density across the nanowire at a high
transverse field (1.04 MV/cm). When the cross section is reduced
from 8 $\times$ 8 nm$^2$ (bottom right panel) to 3 $\times$ 3 nm$^2$
(top left), the onset of volume inversion is evident. The color
bar on the right is in 5 $\times10^{18}$ cm$^{-3}$.} \label{eDenVsCross}
\end{figure}

As the cross section of the SiNW decreases, the channel electrons
are distributed throughout the silicon volume as opposed to just
within a thin channel at the Si/SiO$_2$ interface right below the gate,
as with conventional MOSFETs. The transition from surface inversion
to volume inversion occurs gradually and the cross section at
which the entire silicon is inverted depends on the electron sheet density.\cite{ShishirJCE08}
Fig. \ref{eDenVsCross} shows the variation of the electron density across the wire with varying wire
dimensions (gate is on top). When the cross section is decreased from 8 $\times$ 8
nm$^2$ (bottom right panel) to about 6 $\times$ 6 nm$^2$ (bottom left panel), the onset of volume inversion results in an increase in the average distance of the electrons from the top interface, where the electric field is highest, so SRS is reduced [the second and third terms in the SRS overlap integral (\ref{OverlapSRy}) drop with a decrease of the avarage electronic position from the interface, and are therefore sensitive to volume inversion]. But, once the wire cross section reaches
about 6 $\times$ 6 nm$^2$, the silicon volume is fully inverted, so
further reduction of the cross section simply results in a decrease
of the average distance of the electrons from the interfaces (all four of them), thereby resulting in more surface-roughness scattering.
Consequently, for wires with the cross section smaller
than 6 $\times$ 6 nm$^2$, volume inversion does not offer an
advantage to electronic transport.

\subsection{Mobility Variation with the SiNW Cross Section -- the Big Picture}\label{MobVarVsCS}

Fig. \ref{MuVsCross} shows the variation of the electron mobility when the SiNW cross section is varied from
8 $\times$ 8 nm$^2$ to 3 $\times$ 3 nm$^2$, for a high (red), moderate (blue), and low (green) transverse electric field from the gate.

\begin{figure} [h]
\centering \centering
\includegraphics[width=2.5in]{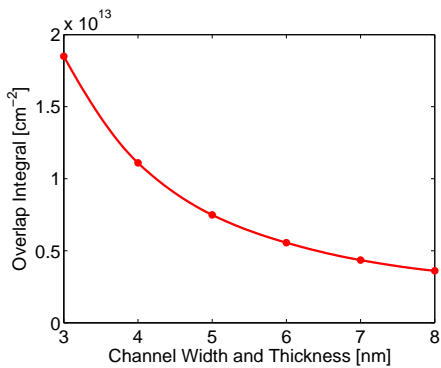}
\caption{  $\quad$(Color online) Variation of the electron-phonon intrasubband scattering overlap with SiNW cross section at a low transverse field (1.4 $\times10^{-2}$ MV/cm).
This results in a fourfold increase in the intrasubband phonon
scattering rates, as the thickness and width of square wires are
varied from 8 nm to 3 nm.} \label{OverlapVsCross}
\end{figure}

\textit{Low and moderate transverse fields}: with decreasing wire cross section, the intrasubband phonon scattering increases due to the increase in the electron-phonon overlap integral (Fig. \ref{OverlapVsCross}); intersubband scattering and intervalley phonon scattering decrease due to subband modulation; SRS increases due to the increase in the first term in the SRS overlap integral with increasing confinement. Overall, the mobility decreases with decreasing wire cross section, very weakly for larger wires and much more rapidly for wires roughly smaller than about 5$\times$5 nm$^2$.
As the wire cross section increases above above 7 $\times$ 7 nm$^2$, we observe a very weak mobility variation that results from the competition between an increase in intersubband scattering (number of occupied subbands increases) and a decrease in intrasubband scattering (electron-phonon overlap integral decreases). A similar weak dependence of the electron mobility with increasing cross section has been reported by Jin {\it et al.} \cite{JinJAP07} for cylindrical SiNWs with diameters greater than 6 nm.

\textit{High transverse fields}: As the wire cross section is reduced from 8 $\times$ 8 nm$^2$ to 5 $\times$ 5 nm$^2$,
the first term in the SRS overlap integral (\ref{OverlapSRy}) increases, whereas the second and the third terms decrease due to the onset of volume inversion (Fig. \ref{eDenVsCross}). Consequently, the mobility shows a very small change for these cross sections. But, when the wire cross section is smaller than 5 $\times$ 5 nm$^2$, the benefits of volume inversion are lost and all the terms in the SRS overlap integral increase with decreasing wire cross section.

\textit{Transverse-field independence of the electron mobility for very thin wires}: We also notice that the transverse-field dependence of the electron mobility weakens with decreasing wire cross section and becomes virtually unimportant for SiNWs thinner than about 5 $\times$ 5 nm$^2$. Irrespective of the effective field from the gate, the mobility decreases monotonically with increasing spatial confinement, with the limiting mechanisms being the steady increase in the field-independent, confinement-induced part of the SRS (\ref{OverlapSRy}) and the increase in intrasubband phonon scattering. Of these  two mechanisms that limit the electron mobility in ultra-thin wires, the SRS scattering dominates.

\section{Conclusion}\label{Concl}
In ultrathin SiNWs, both electrons and acoustic phonons experience 2D confinement. In wires surrounded by SiO$_2$, an acoustically softer material, the acoustic phonon group velocity is lowered to almost half of its bulk silicon value, and leads to enhanced electron--acoustic phonon scattering rates. The electron mobility calculated while accounting for the modification to the acoustic phonon spectrum due to confinement is about 10$\%$ lower than the mobility calculated with bulk acoustic phonons. This result clearly emphasizes the need to account for the acoustic phonon confinement when calculating the electrical properties of SiNWs.

We systematically account for confined acoustic phonons in the calculation and find that the mobility decreases with decreasing wire cross section, very weakly for thicker wires and much more rapidly for wires roughly thinner than about 5$\times$5 nm$^2$. Also, we find that the transverse-field dependence of the electron mobility weakens with decreasing wire cross section and becomes virtually unimportant for SiNWs thinner than about 5 $\times$ 5 nm$^2$.

For thicker wires at low and moderate transverse fields, the slow decrease of mobility with decreasing wire cross section is governed primarily by the interplay between the beneficial subband modulation (less intersubband and intervalley scattering) and the detrimental increase in intrasubband scattering. At higher fields, however, the weak mobility variation with decreasing cross section stems from the competing influences of the beneficial volume inversion and the detrimental enhancement of the confined-induced surface-roughness scattering term.

For very thin wires (below the 5$\times$ 5 nm$^2$ cross section), the mobility decreases monotonically with increasing spatial confinement and becomes virtually independent of the transverse electric field. This occurs primarily due to the increase in the field-independent, confinement-induced part of the SRS (\ref{OverlapSRy}),  and secondly due to the increase in intrasubband phonon scattering. In contrast to bulk MOSFETs, in which the surface roughness scattering plays an important role only for high fields from the gate, electrons in very thin SiNWs are strongly influenced by the roughness regardless of the transverse field. This finding is important both for FETs with multiple gates, such as the FinFET, \cite{Hisamoto00} as well as for  ungated ultrathin wires used for thermoelectric applications \cite{HicksPRB93} or interconnects.

This work has been supported by the National Science Foundation through the University of Wisconsin
MRSEC.

\appendix
\section{Phonon Scattering}\label{phScatDeriv}
\subsection{Bulk Acoustic Phonon Scattering Rate}\label{AcScatDeriv}
The electron wavefunction in subband $n$, taking into account
confinement along the $y$ and $z$ directions and free motion along
$x$, is

\begin{equation}\label{wfn}
\Psi_n(\textbf{r})=\psi_n(y,z)e^{ik_xx}.
\end{equation}

The displacement field $\textbf{u}$ due to longitudinal phonons in
the second-quantization form can be written as

\begin{equation}\label{PhDisp}
\textbf{u}(\textbf{r})=\sum_{\textbf{q}}\sqrt{\frac{\hbar}{2\rho\Omega\omega_\textbf{q}}}
\left( a_\textbf{q}e^{iqr}+a_\textbf{q}^\dag
e^{-iqr}\right)\textbf{e}_\textbf{q},
\end{equation}
where, $a_\textbf{q}$ and $a_\textbf{q}^\dag$ are the phonon
annihilation and creation operators respectively, $\Omega$ is the
volume, $\rho$ is the density, and $\textbf{e}_\textbf{q}$ is the
polarization vector.

The perturbing potential which goes into the matrix element
calculation is given by

\begin{equation}\label{AcPhHam}
\mathrm{H}_{ac}=\Xi_{ac}\nabla\cdot\textbf{u},
\end{equation}
where $\Xi_{ac}$ is the acoustic deformation potential and
$\textbf{u}$ is the phonon displacement given by Eq.(\ref{PhDisp}).

From Eqs. (\ref{PhDisp}) and (\ref{AcPhHam}), we can write the
perturbing potential as
\begin{equation}\label{AcPhHamNew}
\mathrm{H}_{ac}=\Xi_{ac}\sum_{\textbf{q}}\sqrt{\frac{\hbar}{2\rho\Omega\omega_\textbf{q}}}
\textbf{e}_\textbf{q}\cdot{i}\textbf{q}\left(
a_\textbf{q}e^{iqr}-a_\textbf{q}^\dag e^{-iqr}\right).
\end{equation}

\noindent The matrix element after integrating over phonon
coordinates, for scattering from subband $n$ with wavevector
$\textbf{k}_{x}$ to subband $m$ with $\textbf{k}_{x}^{'}$ is given
by

\begin{equation}\label{AcPhMatrix}
\begin{aligned}
\mathrm{M}_{nm}&(\textbf{k}_x,\textbf{k}_x^{'})=\Xi_{ac}\sqrt{\frac{\hbar}{2\rho\Omega\omega_\textbf{q}}}
q\left(N_{\textbf{q}}+\frac{1}{2}{\mp}\frac{1}{2}\right)^{\frac{1}{2}}\\
\times
 &\iint{\left[\psi_n(y,z)e^{i\left(q_yy+q_zz
\right)}\psi_m(y,z)\right]dy\,dz}\\
\times
&\frac{1}{L_x}\int{e^{i\left(k_x-k_x^{'}{\pm}q_x\right)x}{\,}dx},
\end{aligned}
\end{equation}
where $N_{\textbf{q}}$ is the number of phonons given by the
Bose-Einstein distribution function

\begin{equation}\label{BoseNq}
N_{\textbf{q}}=\frac{1}{e^{\frac{\hbar\omega_\textbf{q}}{K_BT}}-1}.
\end{equation}

Defining $\mathcal{I}_{nm}\left(q_y,q_z\right)$ as
\begin{equation}\label{Inm}
\mathcal{I}_{nm}\left(q_y,q_z\right)=\iint{\left[\psi_n(y,z)e^{i\left(q_yy+q_zz
\right)}\psi_m(y,z)\right]dy\,dz}
\end{equation}

\noindent and after integrating over $x$, Eq. (\ref{AcPhMatrix})
yields
\begin{equation}\label{AcPhMatrixSq}
\begin{aligned}
\left|\mathrm{M}_{nm}(\textbf{k}_x,\textbf{k}_x^{'})\right|^{2}&=\Xi_{ac}^{2}\frac{\hbar}
{2\rho\Omega\omega_\textbf{q}}q^{2}\left(N_{\textbf{q}}+\frac{1}{2}{\mp}\frac{1}{2}\right)\\
\times
&\left|\mathcal{I}_{nm}\left(q_y,q_z\right)\right|^{2}\delta(k_x-k_x^{'}{\pm}q_x).
\end{aligned}
\end{equation}

\noindent In the equipartition approximation, the phonon number
becomes
\begin{equation}\label{BoseNqEqPart}
N_{\textbf{q}}{\approx}N_{\textbf{q}}+1{\approx}\frac{K_BT}{\hbar\omega_\textbf{q}},
\end{equation}
so using the equipartition approximation and linear dispersion
relation for acoustic phonons defined by
$\omega_{\textbf{q}}=\upsilon_{s}q$ in Eq. (\ref{AcPhMatrixSq}), we
get
\begin{equation}\label{AcPhMatrixSqNew}
\begin{aligned}
\left|\mathrm{M}_{nm}(\textbf{k}_x,\textbf{k}_x^{'})\right|^{2}&=2\Xi_{ac}^{2}\frac{{\hbar}q^{2}}
{2\rho\Omega\upsilon_{s}q}\frac{K_BT}{\hbar\upsilon_{s}q}\\
\times
&\left|\mathcal{I}_{nm}\left(q_y,q_z\right)\right|^{2}\delta(k_x-k_x^{'}+q_x),
\end{aligned}
\end{equation}
where $\upsilon_{s}$ is the sound velocity in the crystal.

Acoustic phonon scattering rate using Fermi's Golden Rule with the
elastic scattering approximation is given by
\begin{equation}\label{GammaAc}
\begin{aligned}
\Gamma_{nm}^{ac}(\textbf{k}_x)=\frac{2\pi}{\hbar}\sum_{q_{\|},k_x^{'}}\left|\mathrm{M}_{nm}
(\textbf{k}_x,\textbf{k}_x^{'})\right|^{2}\delta(\mathcal{E}-\mathcal{E}^{'}),
\end{aligned}
\end{equation}
where $\mathcal{E}$ and $\mathcal{E'}$ are the initial and final
energies of the scattered electron in the parabolic band
approximation, respectively.

Substituting Eq. (\ref{AcPhMatrixSqNew}) in Eq. (\ref{GammaAc}) and
changing the sum to an integral, we get

\begin{equation}\label{GammaAc1}
\begin{aligned}
\Gamma_{nm}^{ac}(\textbf{k}_x)=&\frac{2\pi\Omega\Xi_{ac}^{2}K_BT}{\rho\Omega\upsilon_{s}^{2}\hbar}
\iint{\left|\mathcal{I}_{nm}\left(q_y,q_z\right)\right|^{2}}\frac{dq_ydq_z}{4\pi^{2}}\\
\times &\int{\delta\left(k_x-k_x^{'}+q_x\right)
\delta(\mathcal{E}-\mathcal{E}^{'})\,\frac{dk_x^{'}}{2\pi}}.
\end{aligned}
\end{equation}

\noindent To evaluate Eq. (\ref{GammaAc1}), let us rewrite it as
\begin{equation}\label{GammaAc2}
\begin{aligned}
\Gamma_{nm}^{ac}(\textbf{k}_x)=\frac{2\pi\Xi_ac^{2}K_BT}{\hbar\rho\upsilon_{s}^{2}}
\mathcal{D}_{nm}\mathcal{I}_2,
\end{aligned}
\end{equation}
where
\begin{equation}\label{I2}
\mathcal{I}_{2}=\frac{1}{2\pi}\int\delta(k_x-k_x^{'}+q_x)\delta(\mathcal{E}-\mathcal{E}^{'})\,dk_x{'},
\end{equation}

\begin{equation}\label{Dnm}
\begin{aligned}
\mathcal{D}_{nm}=\frac{1}{4\pi^{2}}\iint{\left|\mathcal{I}_{nm}\left(q_y,q_z\right)\right|^{2}}dq_ydq_z.
\end{aligned}
\end{equation}

\noindent The above equation can be written as

\begin{equation}\label{Dnm1}
\begin{aligned}
\mathcal{D}_{nm}=&\frac{1}{4\pi^{2}}\int{dq_y}\int{dq_z}\\
\times &\iint\left[\psi_n(y,z)e^{i\left(q_yy+q_zz
\right)}\psi_m(y,z)\right]dy\,dz\\
\times &\iint\left[\psi_n(y^{'},z^{'})e^{-i\left(q_yy^{'}+q_zz^{'}
\right)}\psi_m(y^{'},z^{'})\right]dy^{'}\,dz^{'}.
\end{aligned}
\end{equation}

\noindent Using the identity
\begin{equation}\label{deltaFn}
\begin{aligned}
\frac{1}{2\pi}\int{dq_y}e^{iq_y(y-y^{'})}=\delta(y-y^{'}),
\end{aligned}
\end{equation}
we can write (\ref{Dnm1}) as
\begin{equation}\label{Dnm2}
\mathcal{D}_{nm}=\iint|\psi_n(y,z)|^2|\psi_m(y,z)|^2\,dy\,dz\, .
\end{equation}

Adding a nonparabolicity factor $\alpha$ and converting the
$dk_x^{'}$ integration to $d\mathcal{E'}$ integration, the integral
in Eq. (\ref{I2}) simplifies to
\begin{equation}\label{I2New}
\mathcal{I}_{2}=\frac{1}{2\pi}\sqrt{\frac{m^{*}}{2\hbar^{2}}}\frac{1+2\alpha\mathcal{E}_f}
{\sqrt{\mathcal{E}_f(1+\alpha\mathcal{E}_f)}},
\end{equation}
where $\mathcal{E}_f$ is the final kinetic energy of the electron
after scattering. It is defined in terms of initial parabolic
kinetic energy $\mathcal{E}_{kx}$ as
\begin{equation}\label{AcEf}
\mathcal{E}_f=\mathcal{E}_n-\mathcal{E}_m+\frac{\sqrt{1+4\alpha\mathcal{E}_{kx}}-1}{2\alpha}.
\end{equation}

Substituting the above simplified integrals in Eq. (\ref{GammaAc2})
we get

\begin{equation}\label{GammaAcFinal}
\begin{aligned}
\Gamma_{nm}^{ac}(\textbf{k}_x)=\frac{\Xi_ac^{2}K_BT\sqrt{m^{*}}}{\sqrt{2}\hbar\rho\upsilon_{s}^{2}}
\mathcal{D}_{nm}\frac{1+2\alpha\mathcal{E}_f}{\sqrt{\mathcal{E}_f(1+\alpha\mathcal{E}_f)}}\Theta(\mathcal{E}_f).
\end{aligned}
\end{equation}
where $\Theta(\mathcal{E}_f)$ is the Heaviside step function that
ensures a positive kinetic energy after scattering.

\subsection{Intervalley Scattering Rate}\label{IntValPhDeriv}
Intervalley scattering can be mediated by long-wavevector acoustic
phonons or non-polar optical phonons. Intervalley scattering is
modeled using the non-polar optical phonon model.

The perturbing potential which goes into the matrix element
calculation is given by

\begin{equation}\label{IvPhHam}
\mathrm{H}_{iv}(\textbf{q})=\Xi_{iv}\textbf{e}_\textbf{q}\cdot\textbf{u},
\end{equation}
where $\Xi_{iv}$ is the intervalley deformation potential and
$\textbf{u}$ is the phonon displacement given by Eq. (\ref{PhDisp}).

Assuming intervalley phonons to be dispersionless, \textit{i.e.},
${\omega}_{\textbf{q}}=\omega_{0}$, Eq. (\ref{IvPhHam}) can be
written as

\begin{equation}\label{IvPhHam1}
\mathrm{H}_{iv}=\Xi_{iv}\sum_{\textbf{q}}\sqrt{\frac{\hbar}{2\rho\Omega\omega_{0}}}\left(
a_\textbf{q}e^{iqr}+a_\textbf{q}^\dag e^{-iqr}\right).
\end{equation}

\noindent The matrix element for intervalley scattering is given by

\begin{equation}\label{IvPhMatrix}
\begin{aligned}
\mathrm{M}_{nm}(\textbf{k}_x,\textbf{k}_x^{'})=&\Xi_{iv}\sqrt{\frac{\hbar}{2\rho\Omega\omega_{0}}}
\left(N_{\textbf{q}}+\frac{1}{2}{\mp}\frac{1}{2}\right)^{\frac{1}{2}}\\
\times\ &\mathcal{I}_{nm}\left(q_y,q_z\right)
\delta(k_x-k_x^{'}{\pm}q_x),
\end{aligned}
\end{equation}
where the emission and absorption of an optical phonon results in
$\left(N_{\textbf{q}}+1\right)^{\frac{1}{2}}$ and
$\left(N_{\textbf{q}}\right)^{\frac{1}{2}}$, respectively, and
$\mathcal{I}_{nm}\left(q_y,q_z\right)$ is the overlap integral
defined in Eq. (\ref{Inm}).

Following the procedure outlined above to calculate acoustic phonon
scattering rate and accounting for the inelastic nature of
scattering due to optical phonons, we can write the optical phonon
scattering rate as

\begin{equation}\label{GammaIv}
\begin{aligned}
\Gamma_{nm}^{iv}(\textbf{k}_x)=&\frac{\pi\hbar\Xi_{iv}^{2}}{\rho\omega_0}
\left(N_{\textbf{q}}+\frac{1}{2}{\mp}\frac{1}{2}\right)\\
\times
&\iint{\left|\mathcal{I}_{nm}\left(q_y,q_z\right)\right|^{2}}\frac{dq_ydq_z}{4\pi^{2}}\\
\times &\int{\delta\left(k_x-k_x^{'}{\pm}q_x\right)
\delta(\mathcal{E}-\mathcal{E}^{'}{\pm}\hbar\omega_{0})\,\frac{dk_x^{'}}{2\pi}},
\end{aligned}
\end{equation}
where $\delta(\mathcal{E}-\mathcal{E}^{'}{\pm}\hbar\omega_{0})$
ensures the conservation of energy after absorption (top sign) and
emission (bottom sign) of a phonon of energy $\hbar\omega_{0}$.
Simplifying this using the approach followed in the previous section
(\ref{AcScatDeriv}), intervalley phonon scattering rate can be
written as

\begin{equation}\label{GammaIvFinal}
\begin{aligned}
\Gamma^{iv}_{nm}\left(\textbf{k}_x\right)=\frac{\Xi^2_{iv}\sqrt{m^*}}{2\sqrt{2}\hbar\rho\omega_0}\
&\left(N_{\textbf{q}}+\frac{1}{2}\mp\frac{1}{2}\right)\
\mathcal{D}_{nm} \\
\times
&\frac{(1+2\alpha\mathcal{E}_f)}{\sqrt{\mathcal{E}_f(1+\alpha\mathcal{E}_f)}}\
\Theta(\mathcal{E}_f),
\end{aligned}
\end{equation}
where the final kinetic energy $\mathcal{E}_f$ is similar to that in
Eq. (\ref{AcEf}), with $\pm\hbar\omega_0$ to account for absorption
(top sign) and emission (bottom sign) of a phonon.

\subsection{Confined Acoustic Phonon Scattering Rate}\label{ConfAcPhDeriv}

Using the \emph{xyz} algorithm, the normalized displacement
components for the $J^{th}$ acoustic phonon mode in terms of a
complete set of basis functions {$\Phi_{\lambda}$} can be written as
\begin{equation}\label{DispCompConAcPh}
\begin{aligned}
u_{J,i}=\alpha_{J}\chi_{J,i\lambda}\Phi_{\lambda},
\end{aligned}
\end{equation}
where $i=(x,y,z)$ represents one of the components of the
displacement, $\alpha_{J}$ is the normalization constant and
$\chi_{J,i\lambda}$ are the coefficients of the basis functions.

Taking the center of cross section of the wire as the origin, the
basis functions in terms of powers of Cartesian coordinates in the
lateral directions are

\begin{equation}\label{BasisFn}
\begin{aligned}
\Phi_{\lambda}(x,y,z)=\left(\frac{2z}{W}\right)^{r}\left(\frac{2y}{H}\right)^{s}e^{iq_{x}x},
\end{aligned}
\end{equation}
where $\lambda=(r,s)$, $q_{x}$ is the longitudinal wavevector of the
acoustic phonon mode along the axis of the wire, and $W$ and $H$ are
the width and thickness of the wire, respectively.

Following the normalization procedure indicated in Ref.
\onlinecite{NishiguchiJPCM97}, we get the normalization constant in
Eq. (\ref{DispCompConAcPh}) to be

\begin{equation}\label{alphaJ}
\begin{aligned}
\alpha_{J}=\frac{1}{\sqrt{WHL_x}}\sqrt{\frac{\hbar}{2\omega_{J}}}\frac{1}
{\sqrt{\boldsymbol{\chi}_{J}^{\dag}\mathrm{E}\boldsymbol{\chi}_{J}}},
\end{aligned}
\end{equation}
where $L_x$ is the length of the wire, $\omega_{J}$ is the frequency
of the $J^{th}$ phonon mode, and $\boldsymbol{\chi}_{J}$ is the
eigenvector corresponding to $\omega_{J}$. $\mathrm{E}$ is the
matrix as defined in Ref. \onlinecite{NishiguchiJPCM97}.

The acoustic phonon field, which is used to determine the perturbing
potential, is given by

\begin{equation}\label{DispFieldConfAc}
\begin{aligned}
\textbf{u}=\sum_{J,q_{x}}[a_{Jq_{x}}+a^{\dag}_{Jq_{x}}]\mathbf{e_q}.
\end{aligned}
\end{equation}

Considering Eq. (\ref{DispFieldConfAc}) instead of Eq.
(\ref{PhDisp}) to represent the phonon displacement, the matrix
element given by Eq. (\ref{AcPhMatrix}) can be rewritten as
\begin{equation}\label{ConfAcPhMatrix}
\begin{aligned}
\mathrm{M}_{nm}(k_x,k_x^{'})=&\Xi_{ac}\alpha_{J}
\left(N_{Jq_{x}}+\frac{1}{2}{\mp}\frac{1}{2}\right)^{\frac{1}{2}}\mathcal{L}_{nm}(J,q_{x})\\
\times
&\frac{1}{L_x}\int{e^{i\left(k_x-k_x^{'}{\pm}q_x\right)x}{\,}dx},
\end{aligned}
\end{equation}
where
$N_{Jq_{x}}=\left[{\exp(\hbar\omega_J(q_x)/k_BT)-1}\right]^{-1}$ is
the number of acoustic phonons of energy $\hbar\omega_{J}(q_x)$ and
$\mathcal{L}_{nm}(J,q_{x})$ is the electron-phonon overlap integral
given by
\begin{equation}\label{LnmConfAc}
\begin{aligned}
\mathcal{L}_{nm}&(J,q_{x})=\iint\Bigg[\psi_n(y,z)\Bigg\{\frac{2r\chi_{J,1\lambda}}{W}
\left(\frac{2z}{W}\right)^{r-1}\left(\frac{2y}{H}\right)^{s}\\
&+\frac{2s\chi_{J,2\lambda}}{H}\left(\frac{2z}{W}\right)^{r}\left(\frac{2y}{H}\right)^{s-1}\\
&+iq_{x}\chi_{J,3\lambda}\left(\frac{2z}{W}\right)^{r}\left(\frac{2y}{H}\right)^{s}\Bigg\}
\psi_m(y,z)\Bigg]dy\,dz.
\end{aligned}
\end{equation}

\noindent The square of the matrix element is then given by
\begin{equation}\label{ConfAcPhMatrixSq}
\begin{aligned}
\left|\mathrm{M}_{nm}(k_x,k_x^{'})\right|^{2}=\Xi_{ac}^{2}&
\left(N_{Jq_{x}}+\frac{1}{2}{\mp}\frac{1}{2}\right)\left|\alpha_{J}\right|^{2}\\
\times
&\left|\mathcal{L}_{nm}(J,q_{x})\right|^{2}\delta(k_x-k_x^{'}{\pm}q_x).
\end{aligned}
\end{equation}

Considering the confined acoustic phonon scattering to be inelastic,
the scattering rate can now be written as

\begin{equation}\label{ConfAcPhGamma1}
\begin{aligned}
\Gamma_{nm}^{ac}&\left(k_x\right)=\frac{2\pi\Xi_{ac}^{2}}{\hbar}\sum_{J,q_{x}}
\left(N_{Jq_{x}}+\frac{1}{2}{\mp}\frac{1}{2}\right)\left|\alpha_{J}\right|^{2}\\
\times
&\left|\mathcal{L}_{nm}(J,q_{x})\right|^{2}\delta(k_x-k_x^{'}{\pm}q_x)\delta(E-E^{'}{\pm}\hbar\omega_{J}(q_x)),
\end{aligned}
\end{equation}
where the upper and lower signs denote absorption and emission of an
acoustic phonon of energy $\hbar\omega_{J}(q_x)$, respectively.
Integrating over $q_{x}$ and including the non-parabolicity factor,
the scattering rate can be written as

\begin{equation}\label{ConfAcPhGamma2}
\begin{aligned}
\Gamma_{nm}^{ac}&(k_x)=\frac{\Xi_{ac}^{2}}{2WH}\sum_{J}\int
dq_{x}\left(N_{Jq_{x}}+\frac{1}{2}{\mp}\frac{1}{2}\right)\left|\alpha_{J}\right|^{2}\\
\times
&\left|\mathcal{L}_{nm}(J,q_{x})\right|^{2}\delta(E-E^{'}{\pm}\hbar\omega_{J}(q_x)),
\end{aligned}
\end{equation}
where the total energy of the electron before ($E$) and after ($E'$)
scattering are defined as
\begin{eqnarray}\label{EEdash}
\begin{aligned}
E&=\mathcal{E}_n+\frac{\sqrt{1+4\alpha\frac{\hbar^{2}k_x^{2}}{2m}}-1}{2\alpha},
\\
E^{'}&=\mathcal{E}_m+\frac{\sqrt{1+4\alpha\frac{\hbar^{2}(k_x{\pm}q_{x})^{2}}{2m}}-1}{2\alpha},
\end{aligned}
\end{eqnarray}

\noindent The argument of the delta function in Eq.
(\ref{ConfAcPhGamma2}) can have two roots. Using the identity for
delta functions with multiple roots, the final expression for the
scattering rate is
\begin{equation}\label{ConfAcPhGammaFinal}
\begin{aligned}
\Gamma_{nm}^{ac}&(k_x)=\frac{\Xi_{ac}^{2}}{2WH}\sum_{J}\int
dq_{x}\left(N_{Jq_{x}}+\frac{1}{2}{\mp}\frac{1}{2}\right)\left|\alpha_{J}\right|^{2}\\
\times
&\left|\mathcal{L}_{nm}(J,q_{x})\right|^{2}\bigg[\frac{\delta(q_{x}-q_{x1})}{|g^{'}(q_{x1})|}+\frac{\delta(q_{x}-q_{x2})}{|g^{'}(q_{x2})|}\bigg],
\end{aligned}
\end{equation}
where $g(q_{x})=(E-E^{'}{\mp}\hbar\omega_{J}(qx)$), $q_{x1}$ and
$q_{x2}$ are the two possible roots of $g(q_{x})=0$, and
$g^{'}(q_{x1})$ and $g^{'}(q_{x2})$ are the derivatives of
$g(q_{x})$ with respect $q_{x}$ evaluated at $q_{x1}$ and $q_{x2}$,
respectively.

\section{Surface Roughness Scattering}\label{SRSDeriv}
In a very simple model used to describe the surface roughness
scattering, the perturbing potential for the interface normal to the
$y$ direction is given by

\begin{equation}\label{PertPotSRS}
\begin{aligned}
\mathrm{H}_{sr} = e\varepsilon_y(y,z)\triangle(x),
\end{aligned}
\end{equation}
where $\triangle(x)$ is a random function which describes the
deviation of the actual interface from the ideal flat interface and
$\varepsilon_y(y,z)$ is the field normal to the interface. The
scattering matrix calculated using this perturbing potential is

\begin{equation}\label{SRSMatrix}
\begin{aligned}
\mathrm{M}_{nm}(\textbf{k}_x,\textbf{k}_x^{'})=&e\iint{\left[\psi_n(y,z)\varepsilon_y(y,z)\psi_m(y,z)\right]dy\,dz}\\
\times &\frac{1}{L_x}\int{\triangle(x)e^{i(k_x-k_x^{'})x}{\,}dx}.
\end{aligned}
\end{equation}

\noindent Defining the term in the double integral of the above
equation as $\mathcal{F}_{nm}$, the square of the matrix element can
be written as

\begin{equation}\label{SRSMatrixSq}
\begin{aligned}
\left|\mathrm{M}_{nm}(\textbf{k}_x,\textbf{k}_x^{'})\right|^{2}=&e^{2}\mathcal{F}_{nm}^{2}
\frac{1}{L_x^{2}}\int{dx^{'}}\\
\times
&\int{dx}\triangle(x)\triangle(x^{'})e^{i(x-x^{'}){\cdot}(k_x-k_x^{'})}.
\end{aligned}
\end{equation}

\noindent The average value of the matrix element given by Eq.
(\ref{SRSMatrixSq}) over many samples is actually used to calculate
the SRS. The expectation value of the square of the matrix element
is given by

\begin{equation}\label{SRSMatrixSqAv}
\begin{aligned}
\left\langle\left|\mathrm{M}_{nm}(\textbf{k}_x,\textbf{k}_x^{'})\right|^{2}\right\rangle=&e^{2}\mathcal{F}_{nm}^{2}
\frac{1}{L_x^{2}}\int{dx^{'}}\\
\times
&\int{dx}\langle\triangle(x)\triangle(x^{'}){\rangle}e^{i(x-x^{'}){\cdot}q_x},
\end{aligned}
\end{equation}
where the correlation function
$R(x-x^{'})=\langle\triangle(x)\triangle(x^{'})\rangle$ depends only
on the distance $|x-x^{'}|$, and $q_x=(k_x-k_x^{'})$. Redefining
$(x-x^{'})$ as $x^"$ we get

\begin{equation}\label{SRSMatrixSqAv}
\begin{aligned}
\left\langle\left|\mathrm{M}_{nm}(\textbf{k}_x,\textbf{k}_x^{'})\right|^{2}\right\rangle=e^{2}\mathcal{F}_{nm}^{2}
\frac{1}{L_x^{2}}\int{dx^{'}}\int{dx^{"}}R(x^{"})e^{ix^{"}q_x}.
\end{aligned}
\end{equation}
Assuming exponentially correlated surface roughness
\cite{GoodnickPRB85} defined by
$R(x)=\Delta^{2}e^{-\frac{\sqrt{2}|x|}{\lambda}}$ in Eq.
(\ref{SRSMatrixSqAv}), we get

\begin{equation}\label{SRSMatrixSqAv1}
\begin{aligned}
\left\langle\left|\mathrm{M}_{nm}(\textbf{k}_x,\textbf{k}_x^{'})\right|^{2}\right\rangle=e^{2}\mathcal{F}_{nm}^{2}
\frac{1}{L_x^{2}}\Delta^{2}L_x\frac{2\sqrt{2}\lambda}{(q_x^{\pm})^2\lambda^2+2}.
\end{aligned}
\end{equation}

The scattering rate due to the interface imperfections can now be
written as

\begin{equation}\label{GammaSR}
\begin{aligned}
\Gamma_{nm}^{sr}(\textbf{k}_x)=\frac{2\pi}{\hbar}\sum_{\textbf{k}_x^{'}}e^{2}\mathcal{F}_{nm}^{2}
\frac{1}{L_x}\Delta^{2}\frac{2\sqrt{2}\lambda}{q_x^2\lambda^2+2}\delta(\mathcal{E}^{'}-\mathcal{E}).
\end{aligned}
\end{equation}

Converting the sum over $k_x^{'}$ to integral over $dk_x^{'}$, we
get

\begin{equation}\label{GammaSR1}
\begin{aligned}
\Gamma_{nm}^{sr}(\textbf{k}_x)=&\frac{2\pi}{\hbar}e^{2}\mathcal{F}_{nm}^{2}
\frac{2\sqrt{2}\lambda\Delta^{2}}{L_x}\\
\times
&\frac{L_x}{2\pi}\int{dk_x^{'}}\frac{1}{(k_x{\pm}k_x^{'})^2\lambda^2+2}\delta(\mathcal{E}^{'}-\mathcal{E}).
\end{aligned}
\end{equation}

Defining
$\mathcal{E}_i=\frac{\sqrt{1+4\alpha\mathcal{E}_{kx}}-1}{2\alpha}$
and redefining $(q_x^{\pm})^{2}$ as
$2m/\hbar^{2}(\sqrt{\mathcal{E}_i(1+\alpha{\mathcal{E}_i})}\pm
\sqrt{\mathcal{E}_f(1+\alpha\mathcal{E}_f)})^{2}$, the final
expression for surface roughness scattering assuming non parabolic
bands can be written as

\begin{equation}\label{GammaSRFinal}
\begin{aligned}
\Gamma^{sr}_{nm}(k_x,\pm)=\frac{2\sqrt{m^*}e^2}{\hbar^2}&\frac{\Delta^2\Lambda}{2+(q^{\pm}_x)^2\Lambda^2}|\mathcal{F}_{nm}|^2
\\
\times
&\frac{(1+2\alpha\mathcal{E}_f)}{\sqrt{\mathcal{E}_f(1+\alpha\mathcal{E}_f)}}\
\Theta(\mathcal{E}_f).
\end{aligned}
\end{equation}

Unlike the approach detailed above, Ando's model \cite{AndoRMP82} of
interface roughness scattering accounts for deformation of both the
wavefunction and potential due to the imperfection at the Si-SiO$_2$
interfaces. The matrix element including the perturbation to
wavefunction and the potential using Ando's model is given by
\begin{equation}\label{SRSMatAndo}
\begin{aligned}
\mathrm{M}_{nm}(\textbf{k}_x,\textbf{k}_x^{'})=&\iint{dydz}\int{dx}\frac{e^{-ik^{'}_{x}x}}{\sqrt{L}}\bigg\{
\psi_m(y+\triangle{x},z)\\
\times
 &\big[\mathrm{H}_0+\triangle{V}(y+\triangle{x},z)\big]\psi_n(y+\triangle{x},z)\\
&-
\psi_m(y,z)\mathrm{H}_0\psi_n(y,z)\bigg\}\frac{e^{ik_{x}x}}{\sqrt{L}},
\end{aligned}
\end{equation}
where the unperturbed system is represented by the Hamiltonian
$\mathrm{H}_0$ and wavefunction $\psi_n(y,z)$ and the perturbed
system's Hamiltonian and wavefunction are
$\big[\mathrm{H}_0+\triangle{V}(y+\triangle{x},z)\big]$ and
$\psi_m(y+\triangle{x},z)$. Calculating the scattering rate from
this matrix element, we find that the final expression for the SRS
is same as before except that, we have additional wavefunction
deformation terms in the SRS overlap integral $\mathcal{F}_{nm}$.
The SRS overlap integral is given in Eq. (\ref{OverlapSRy}).



\end{document}